# Magnetic SAW RFID sensor based on Love wave for detection of magnetic field and temperature

Prince Mengue, Laurine Meistersheim, Sami Hage-Ali, Cécile Floer, Sébastien Petit-Watelot, Daniel Lacour, Michel Hehn and Omar Elmazria.

Université de Lorraine - CNRS, Institut Jean Lamour (UMR 7198), F-54000, Nancy – France
omar.elmazria@univ-lorraine.fr

*Abstract*— **Magnetic field measurement including a temperature compensation is essential for a magnetic field sensor. This study investigates a magnetic surface acoustic wave (MSAW) sensor in a reflective delay line configuration with two acoustic propagation paths with and without magnetic field sensitive layer. The delay in path with sensitive layer leads to magnetic field detection and the one without enable temperature measurement and thus compensation for the first path. The developed sensor is based on a ZnO/LiNbO₃ Y-cut (X-direction) layered structure as Love wave platform. Love wave as a shear wave being more favorable for magnetic detection. Co-Fe-B is considered as sensitive layer to detect magnetic field changes and is deposited on the top of ZnO, but only on one of the two paths. We combined an original configuration of connected IDTs with a high electromechanical coupling coefficient ($K^2$) mode to improve the signal amplitude. The achieved sensor exhibits a high temperature and magnetic field sensitivity of -63 ppm/°C and -781 ppm/mT, respectively. The temperature compensation method for magnetic field measurement is demonstrated using a differential measurement by subtracting the delay times obtained for the two paths with and without the sensitive layer. Finally, The sensor exhibited good repeatability at various temperatures. Moreover, the device developed allows in addition to the multisensor functionality, the radio frequency identification (RFID) which is necessary for the deployment of sensor networks.**

## I. INTRODUCTION

Magnetic field surface acoustic waves (MSAW) devices have made tremendous progress in various areas in recent years. Compared to other field effect based technologies such as SQUID sensors, Hall-effect sensors or magnetoresistive sensors, the use of magnetoelastic coupling in acoustic wave devices can lead to cost and size reductions, to fully passive devices or even to wireless measurements when the device is connected to antennas [1]. The SAW technology is based on metal interdigital transducers (IDTs) placed over a piezoelectric substrate. It is widely used for sensing applications [1, 2] since the wave can be sensitive to physical and chemical stimuli such as temperature, pressure, deformation, vibration, chemical species etc. This results in a velocity change and thus in a frequency variation of the device. Equation (1) shows the relation between the resonance frequency ($f_r$), the wave velocity (v) and the wavelength (λ) fixed by the spatial period of the IDTs.

$$f_r = \frac{v}{\lambda} \qquad (1)$$

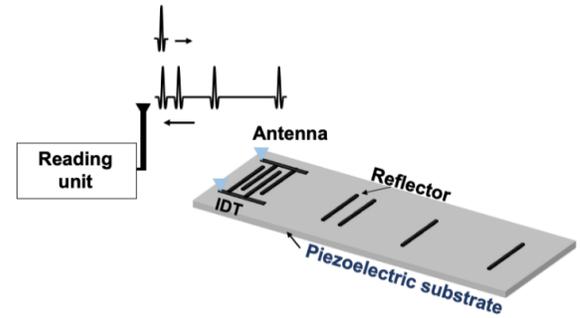

Fig. 1. Interrogation principle of a wireless SAW ID-Tag system

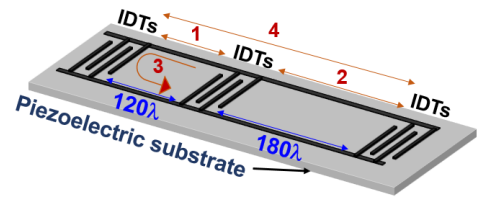

Fig. 2. RFID system with three electrically connected IDTs used in this work

SAW devices can advantageously be combined with a Radio Frequency Identification (RFID) system to be located within a sensor network. The general principle of SAW RFID tags is depicted in Fig. 1. The chip receives a radio frequency signal (RF) through the antenna. This signal is then converted by the IDTs into a surface acoustic wave thanks to the inverse piezoelectric effect. The wave propagates at the surface of the

This work was supported by ANR JCJC SAWGOOD (ANR-18-CE42-0004-01) and ANR PRCE WISTITWIN (ANR-20-CE42-0009-01).
Experiments are carried out at MiNaLor clean-room platform which is partially supported by FEDER and Grand Est Region through the RaNGE project and at IJL Project TUBE-Davm equipment funded by the French PIA project "Lorraine Université d'Excellence" (ANR-15-IDEX-04-LUE).

substrate and is reflected by different reflectors. The reflected signal is then converted back into a RF signal. The position of the reflectors generates echoes in the form of bar codes. Each echo gives a unique identification of the sensor. In order to improve the signal amplitude while keeping an identification feature, a configuration with connected IDTs [4], as illustrated in Fig. 2.

MSAW sensors have a sensitive thin film material placed on the propagation path of the wave. Numerous research works with different magnetic materials such as Nickel (Ni) [5], Galfenol (FeGa) [6], FeCoSiB ([7], [8]), FeCo [9] or CoFeB [1] have already been published. In the presence of a magnetic field, the sensitive magnetoelastic layer undergoes a change in its elastic constants due to the so called "ΔE/ΔG" effect. This create a phase velocity change for the acoustic wave, and thus also a frequency change in the case of a resonator (see equation 1) or a phase/delay change in the case of a delay line as shown in Fig. 3.

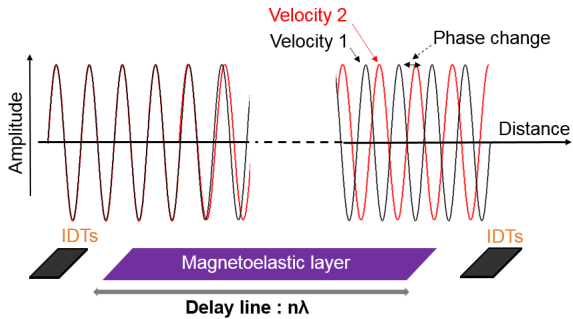

Fig. 3.  Phase change under external magnetic field in a delay line

The conventional configurations, namely the resonator [10] [1] and the regular 2-port delay line ([8], [11], [12]) make up for the majority of the MSAW literature. On the other hand, the ID-tag configuration is still very underdeveloped in this state of the art. However, Li *et al.* showed in 2015 the attractiveness of this configuration [13]. Therefore, the identification function of MSAW sensors remains an important topic to be explored.

Numerous applications also require a highly sensitive magnetic sensor. However, MSAW devices are often affected by the external temperature [14]. Indeed, a thermal expansion and a change in the elastic properties of the structure [15] occur, resulting in a velocity change and thus in a less accurate magnetic field measurement. Therefore, MSAW sensors with a zero temperature coefficient of frequency (TCF) have attracted the attention. Mishra *et al.* carried out studies with single port MSAW resonator based on a CoFeB/ZnO/quartz structure [16]. The structure shows good temperature compensation but the magnetic sensitivity obtained with the Love wave remains relatively low (-35ppm/mT). In [17], Mishra *et al.* demonstrated the additional stabilization of the magnetic properties of the sensitive material in MSAW sensors, using a micro-structuration. Yang *et al.* reached a higher sensitivity of 623 ppm/mT with a MSAW sensor working at 433MHz and based on a CoFeB/SiO$_2$/ZnO/quartz structure [1]. In addition to being temperature compensated, this structure has the advantage of having two sensitive modes. The Love mode (temperature compensated) is sensitive to the magnetic field, and the Rayleigh mode is mainly sensitive to the temperature. However, the Rayleigh mode also exhibits a slight sensitivity to the magnetic field, which is troublesome for accurate temperature detection. Thus, developing a MSAW sensor that performs highly sensitive measurements of both temperature and magnetic field remains a real challenge.

In this manuscript, we report a multifunctional temperature and magnetic field SAW sensor. It also includes a temperature compensation feature for the magnetic field measurement. We used a reflective delay line (R-DL) configuration with connected IDTs to enable the radio frequency identification function (Fig. 2). The FEM modeling used to optimize the Love wave is first explained. Then, the experimental fabrication is presented, followed by the temperature and the magnetic field characterizations. Finally, the temperature compensation of the magnetic field detection is demonstrated.

## II. SIMULATION RESULTS

A two-dimensional finite element method (2D-FEM) model was performed with COMSOL Multiphysics. The structure was simulated in cross-section and a perfectly matched layer (PML) condition was applied at the boundaries to artificially absorb the waves reflected from the edges of the device (Fig.4a). The wavelength (λ) and the metallization ratio were set to 9.2µm and 50%, respectively. Using a Y-cut LiNbO$_3$ substrate with a X-direction propagation, three types of waves are generated, namely the Rayleigh wave, the Love wave and bulk waves [18]. ZnO was used to achieve a high coupling coefficient Love mode in the finale structure and CoFeB was chosen as a magnetic sensitive layer. The physical constants of LiNbO$_3$, ZnO and CoFeB were taken from the literature and are summarized in Table I.

TABLE I
SUMMARY OF THE PHYSICAL CONSTANTS USED IN THE SIMULATIONS

| **LiNbO$_3$** | **ZnO**[19] | **CoFeB**[15] |
|---|---|---|
| Density : 4700 (kg/m$^3$) | Density : 5680 (kg/m$^3$) | Density : 8000 (kg/m$^3$) |
| Elastic constant (GPa) C11: 202.897 C12: 529.177 C13: 749.098 C33: 243.075 C44: 599.034 C66: 748.772 | Elastic constant (GPa) C11: 209.14 C12: 121.14 C13: 105.359 C33: 211.194 C44: 423.729 C66: 442.478 | Elastic constant (GPa) C11: 257 C12: 162 C33: 105 |
| Piezoelectric constant (C/m$^2$) *e15*: 3.69594 *e16*: -2.53384 *e31*: 0.193644 *e33*: 1.30863 | Piezoelectric constant (C/m$^2$) *e15*: -0.48 *e31*: -0.56 *e33*: 1.32 | |
| Relative permittivity *ε11*: 43.6 *ε33*: 29.16 | Relative permittivity *ε11*: 8.54 *ε33*: 10.204 | |

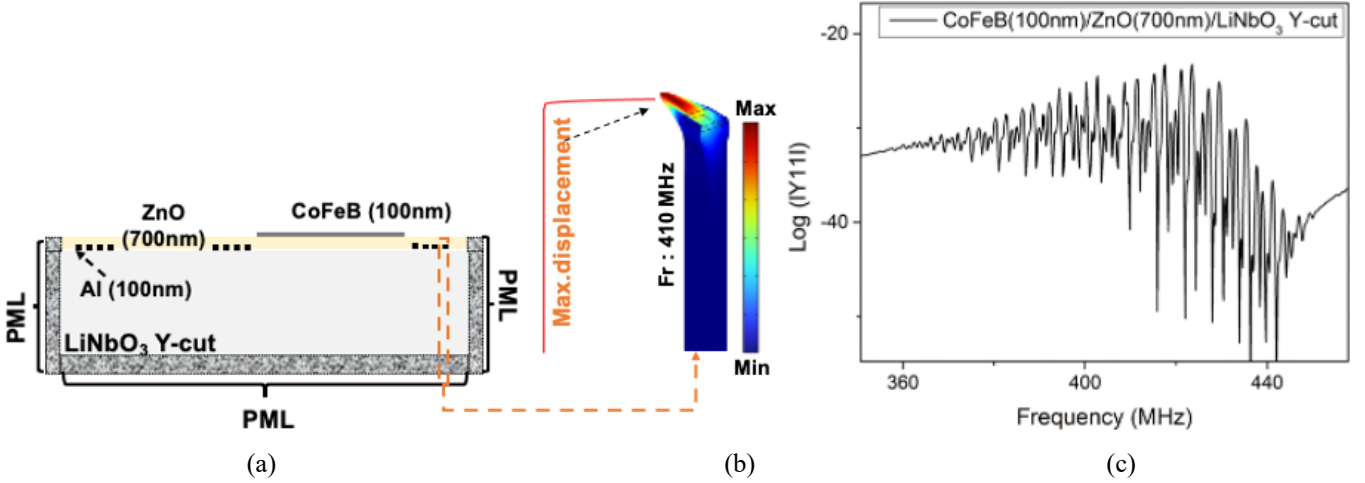

(a) (b) (c)

Fig. 4. (a) 2D schematic view of the simulated MSAW structure, (b) 3D particle displacement for the Love wave at its resonance frequency and (c) 2D simulated admittance response of the CoFeB(100nm)/ZnO(700nm)/LiNbO₃ structure

The thickness of the CoFeB (100 nm) and ZnO (700 nm) were chosen to optimize the Love mode, with a maximum wave displacement at the surface. This allows to have a large proportion of the wave propagating in the magnetic layer and therefore to maximize the sensitivity (Fig. 4b). The simulated admittance response is given in Fig. 4c and shows an operating frequency around 410 MHz.

## III. DESIGN, MATERIALS AND FABRICATION

The structure based on connected IDTs was then manufactured (see schematic Fig 5a, and the top view of fabricated MSAW device Fig 5b). The following technological steps were used: 100 nm-thick aluminum IDTs were patterned with an optical lithography process on the Y-cut LiNbO₃ substrate (X-direction). Design parameters are summarized in Table II. Then, a 700 nm-thick ZnO thin film was deposited over the entire structure by RF magnetron sputtering using the parameters given in [19]. Finally, a 100 nm-thick magnetic CoFeB layer was sputtered along path 2 (180λ) to allow the magnetic field detection, while path 1 (120λ) was protected by masking. Temperature measurements will therefore be possible with a wave propagating in path 1.

TABLE II
CHARACTERISTIC VALUES OF THE FABRICATED DEVICE

| | |
|---|---|
| LiNbO₃ thickness | 0.5 mm |
| Wavelength (λ) | 9.2 µm |
| Metallization ratio | 50 % |
| Number of IDTs finger pairs | 11 |
| Number of reflector electrodes | 200 |
| Al electrode thickness | 100 nm |
| Path 1 | 120λ |
| Path 2 | 180λ |

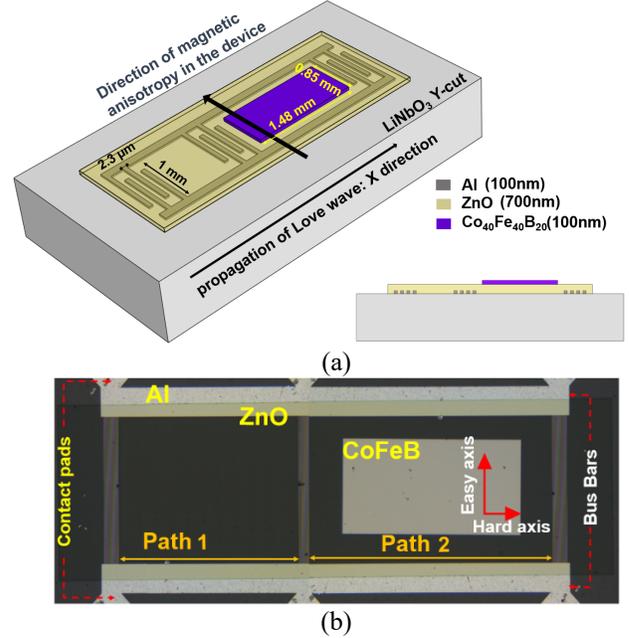

(a)

(b)

Fig. 5. (a) 3D image and cross-sectional view of the fabricated device (b) top view of the manufactured MSAW reflective delay-line

## IV. EXPERIMENTAL RESULTS AND DISCUSSION

### A. Electrical characterization of the MSAW sensor

The electrical characterization of the final device was performed with a RF probe station (Süss Microtec, PM5) and a vector network analyzer (Rohde&Schwarz, ZNL6). Fig. 6a shows the measured $S_{11}$ reflection coefficient for the CoFeB/ZnO/LiNbO₃ Y-cut final structure. A good agreement is found between the experimental and theoretical curves. The fairly wideband oscillations around the Love wave resonance frequency (410 MHz) are typical of delay lines. The time domain response is shown in Fig. 6b after applying the inverse fast Fourier transform (IFFT) to the previous $S_{11}$ curve. The peaks correspond to the different propagation paths of the Love wave inside the structure (previously shown in Figs. 2&5). Peaks 1, 2 and 4 are related to a direct transmission between 2 connected IDTs and peak 3 corresponds to a reflection in the shorter gap (120λ). In this work, we focused only on peaks 1 and 2. A high level (-18 dB) is achieved for peak 1 which is propagating in the gap (120λ) without the magnetic layer. Peak 2 also shows a relatively high level (-24 dB) even though the wave mainly propagates in the CoFeB layer. This result is due to the combination of a multilayer structure with a high electromechanical coupling coefficient ($K^2$) and the use of a direct transmission in the connected IDTs configuration.

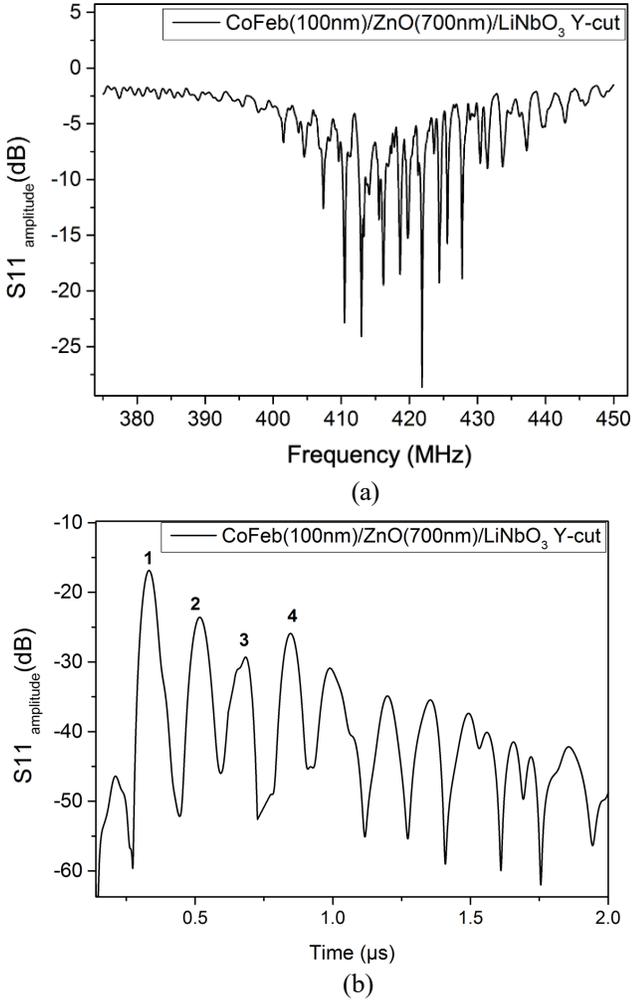

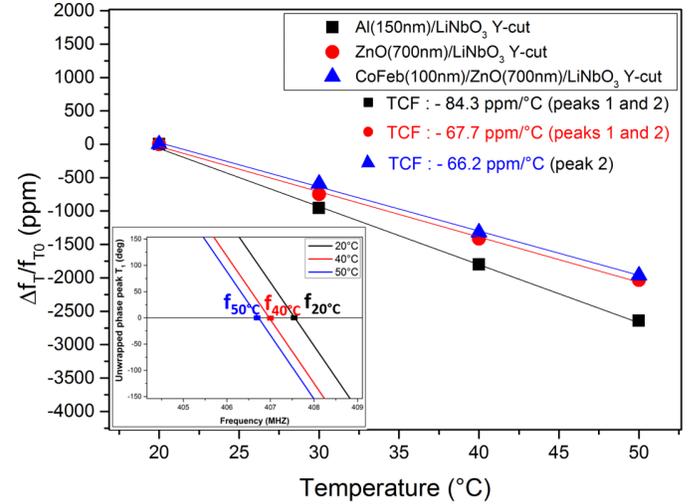

temperature coefficient of frequency (TCF) is -84 ppm/°C and -67.7 ppm/°C for both peaks on the $LiNbO_3$ Y-cut and $ZnO/LiNbO_3$ Y-cut structure, respectively. After the CoFeB deposition, peak 1 has an unchanged sensitivity of -67.7ppm/°C. This is consistent with the previous measurement as there is no magnetic layer in the first gap (120λ). Peak 2 has roughly the same sensitivity to the temperature, namely -66.2 ppm/°C. This proves that the magnetic layer has little influence on the TCF value. The very small TCF difference of the two peaks appears as a promising result that favors the very simple realization of the temperature compensation of peak 2.

Fig. 7. Relative frequency shifts versus temperature for peak 1 and 2 at different manufacturing steps (inset : S11(f) unwrapped phase tracking in the frequency domain)

### C. Magnetic field sensitivity of the MSAW sensor

The final MSAW sensor was investigated, in vacuum, under the magnetic field and the temperature influence. The experimental setup is shown in Fig. 8. The electromagnet provides an in-plane magnetic field. The probing area is placed in a vacuum chamber and the GS probe allows the contact with the device. The vector network analyzer (VNA) monitors the MSAW response. The thermal chuck, consisting of a heating element coupled to a liquid nitrogen, ensures a fair thermal stability during measurements. The vacuum of $10^{-4}$ mbar is also maintained to prevent the elements in the chamber from being damaged during the heating, for example by oxidation.

Fig. 6. Experimental reflection coefficient $S_{11}$ of the $CoFeB/ZnO/LiNbO_3$ structure in (a) the 370-450 MHz frequency range and (b) in the time domain

### B. Temperature sensitivity of the MSAW sensor

For the temperature characterization, the previous system associated with a thermal chuck was used. The frequencies linked to peaks 1 and 2 were tracked with temperature ranging from 25°C to 50°C. These frequencies were obtained using the following method : in the time domain, a time gating was applied to either peak 1 or peak 2. After a Fourrier transform, the frequencies corresponding the zero value of the unwrapped phase of S11(f) were tracked for various temperatures.

The frequency shift evolutions versus the temperature is shown in Fig. 7 at various steps of the fabrication process. The

A similar procedure to the previous one was used for the peak tracking under various magnetic field amplitudes. The magnetization curves of the 100 nm thick CoFeB magnetic layer are given in our previous work [7]. The temperature was first set at 21°C and the magnetic field varied between -4 and 4 mT. Figure 9 shows the relative frequency shifts for peak 1 (red curve) and peak 2 (black curve) with a magnetic field applied along the easy axis (i.e. the direction of preferred magnetization). Peak 1 shows a small deviation due to small temperature variations around 21°C. Peak 2 shows a linear behavior from -0.19mT to 0.69 mT with a high magnetic sensitivity of -781 ppm/mT (291 Hz/µT). This value is higher than the results obtained in the last studies with the same CoFeB thickness on other substrates [11,12]. Moreover, the behavior of peak 1 shows a sensitivity to the magnetic field close to zero. These interesting results make this MSAW device a multifunctional sensor where peaks 1 and 2 can be used for temperature and magnetic field measurements, respectively.

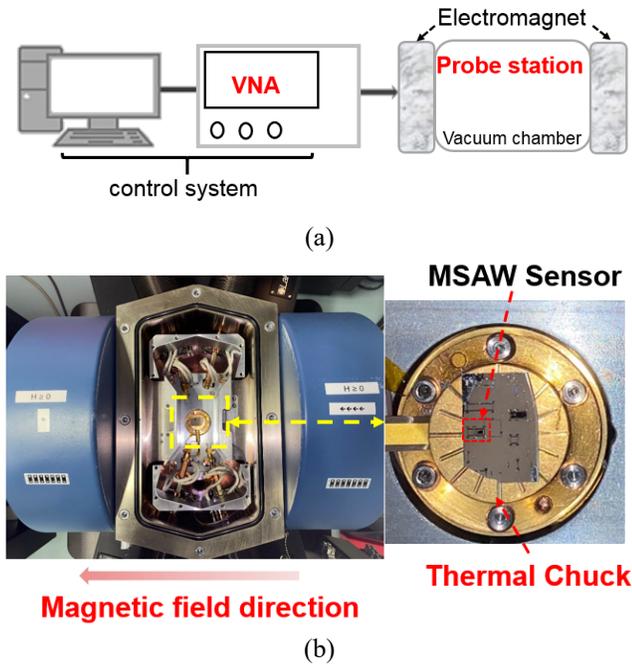

Fig. 8. (a) Schematic layout of the magnetic setup (b) Experimental probe station with an MSAW sensor on the thermal chuck.

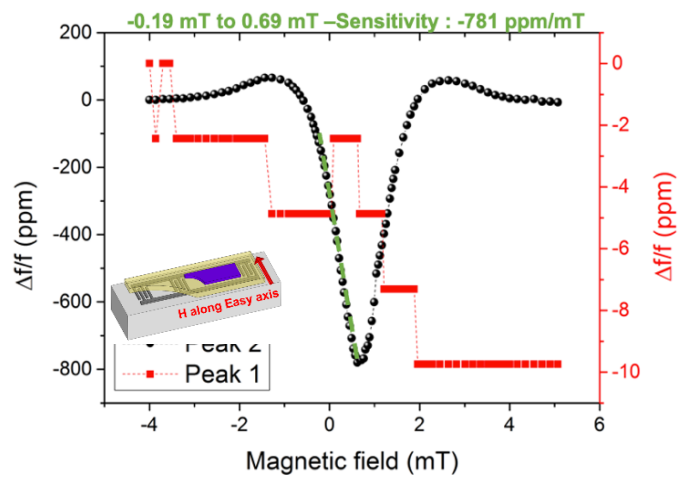

Fig. 9. Relative frequency shifts versus magnetic field (along the easy axis) for peaks 1 and 2 in the CoFeB/ZnO/LiNbO3 Y-cut structure at a fixed 21°C temperature.

*D. Temperature compensation of the magnetic field sensitivity of the MSAW sensor*

As peak 2 is sensitive to the magnetic field and to the temperature, it is required to achieve a temperature compensation in order to obtain a more accurate magnetic field measurement. Fig. 10 shows the magnetic field measurements at different temperatures from 7°C to 37°C for peak 1 (Fig. 10a) and peak 2 (Fig. 10b). Even though the chuck temperature is regulated, the temperature at the sensor surface may slightly fluctuate depending on the experimental conditions. Indeed, the temperature sensor ensuring the regulation is located inside the chuck. The measurements show that a good stability is achieved when the target temperature is close to the room temperature (21 °C). However, for targeted temperatures away from the ambient, the experimental setup showed some limitations in its stabilization. For the measurements at 7°C and 17°C, the temperature drops slightly and continuously, which explains the slight frequency increase (Fig. 10a). For measurements above the ambient (27°C and 37°C), the temperature slightly increases, leading to a frequency decrease (Fig. 10a). These slight temperature variations are obvious for peak 1 as it only responds to a temperature stimuli, whereas they are less visible for peak 2 as its magnetic sensitivity is much more pronounced that the temperature sensitivity. Note that these temperature variations are not a problem in temperature compensation. The measured TCF for peak 1 and 2 being negatives, the frequencies decrease with the increasing temperature.

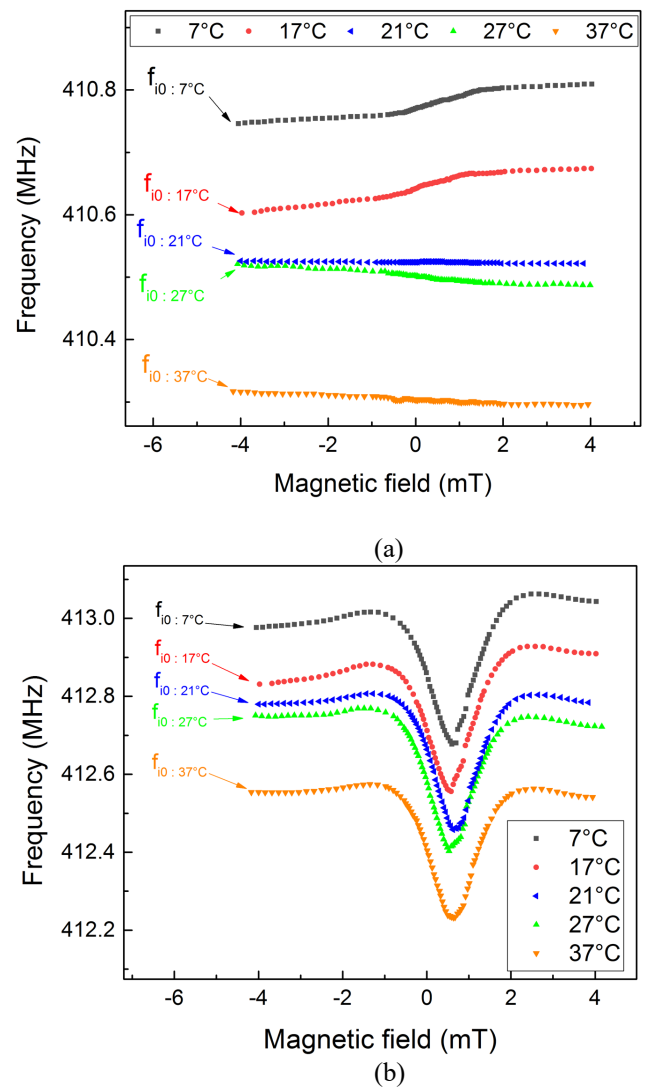

Fig. 10. Relative frequency shifts versus magnetic field (along the easy axis) at different temperature for (a) peak 1 and (b) peak 2 in the final MSAW sensor

Based on the previous measurements (Fig. 10), the temperature compensation of peak 2 was obtained by calculation through the following steps: first, the relative frequency change versus magnetic field was plotted for both peaks (Fig. 11a and 11b). Peak 2 (Fig. 11b) includes both the magnetic field effect illustrated with the characteristic shape of its response and the temperature effect through an unperfect overlap of the curves ($TCF_2 = -66.2$ ppm/°C). Since both peaks have roughly the same temperature sensitivity, the temperature variations subjected by peak 1 are almost the same for peak 2. Therefore, to achieve a magnetic sensitivity of the device with a temperature compensation, the curves from peak 1 have been subtracted from peak 2 responses according to the following relation (2):

$$\left(\frac{\Delta f_i}{f_{i0}}\right)_2 - \frac{TCF_2}{TCF_1}\left(\frac{\Delta f_i}{f_{i0}}\right)_1 \qquad (2)$$

where $TCF_1$, $TCF_2$, $(\Delta f_i/f_{i0})_1$, and $(\Delta f_i/f_{i0})_2$ are the temperature sensitivities and the relative frequency change for peaks 1 and 2, respectively. $\Delta f_i$ is given by $f_i - f_{i0}$, where $f_{i0}$ is the initial frequency at -4mT and $f_i$ is the corresponding frequency in the magnetic field range from -4mT to 4mT.

The results are shown in Fig. 11c and correspond to the temperature compensation of peak 2. A near perfect superposition of the different magnetic sensitivity curves, especially in the range of interest, is visible: this demonstrates the interest of the structure and the associated temperature compensation scheme for the MSAW measurement of the magnetic field under various temperature conditions.

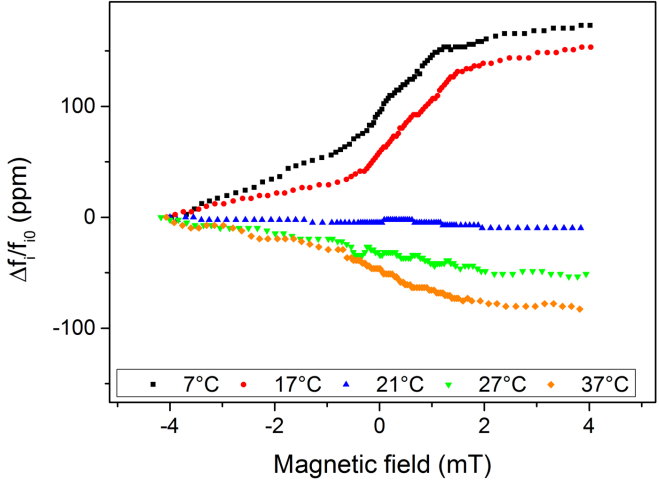

(a)

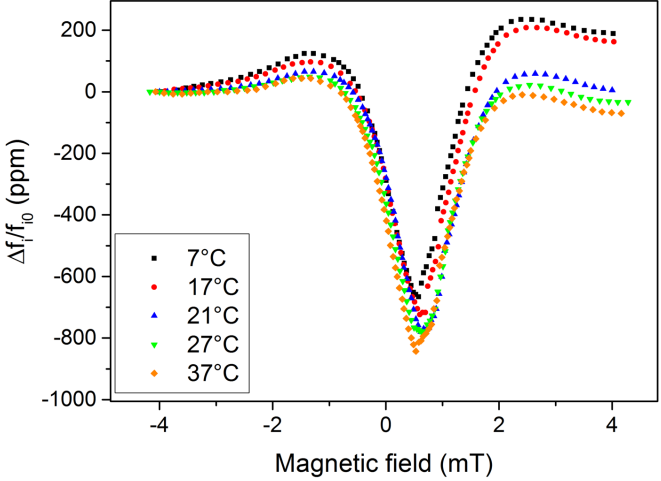

(b)

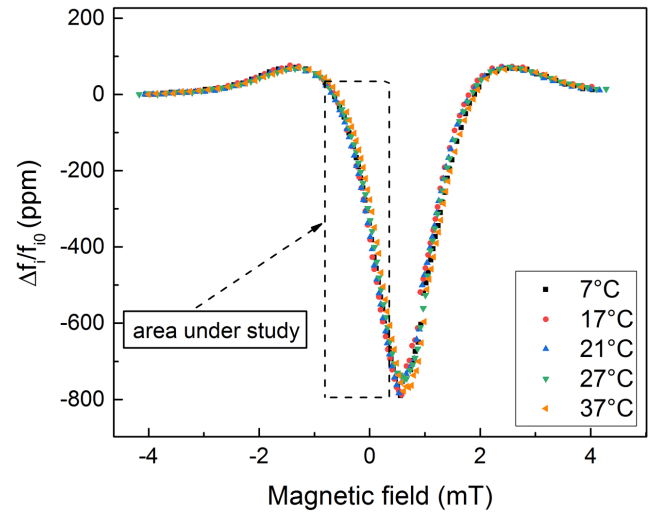

(c)

Fig. 11. Relative frequency shifts versus magnetic field (along the easy axis) at different temperature for (a) peak 1, (b) peak 2 and (c) for peak 2 after temperature compensation.

## V. Conclusion

In this work, we developed a MSAW sensor based on a connected IDTs configuration for measuring both the temperature and the magnetic field amplitude. We used a CoFeB/ZnO/LiNbO$_3$ Y-cut (X-direction) multilayer structure to improve the magnetic field sensitivity and achieve a temperature compensation.

Two peaks, corresponding to acoustic paths without and with a magnetic field sensitive layer, were studied under the influence of temperature and magnetic field. Peak 1 showed a temperature sensitivity of -67.7 ppm/°C and a magnetic sensitivity close to 0. Peak 2 showed a temperature sensitivity of -66.2 ppm/°C and a high magnetic sensitivity of -781 ppm/mT. Hence, we were able to compensate the temperature effect of peak 2 by subtracting the two responses.

These very interesting results introduce the device as a multifunctional MSAW sensor with temperature and magnetic field measurements. In addition, the high signal level allows for greater design flexibility to interrogate the device remotely and thus to extend its potential field of application.

## Acknowledgment

The authors would like to thank Laurent Badie at MiNaLor platform at Institut Jean Lamour for his help with the microfabrication.